\documentclass[doublecol]{epl2} 
\title{Investigation of $^{74}$As decay branching ratio dependence on the host material}
\shorttitle{Investigation of $^{74}$As decay branching ratio dependence on the host material} %Insert here a short version of the title if it exceeds 70 characters

\author{Gy.~Gy\"urky\inst{1} \and J. Farkas\inst{1} \and C. Yal\c c\i n\inst{1,2} \and G.G. Kiss\inst{1} \and Z. Elekes\inst{1} \and Zs.~F\"ul\"op\inst{1} \and E. Somorjai\inst{1}}
\shortauthor{Gy.~Gy\"urky \etal}

\institute{                    
  \inst{1} Institute of Nuclear Research (ATOMKI), H-4001 Debrecen, POB.51., Hungary\\
  \inst{2} Kocaeli University, Department of Physics, TR-41380 Umuttepe, Kocaeli, Turkey
}
\pacs{21.10.Tg}{Lifetimes, widths}
\pacs{23.40.-s}{$\beta$ decay; double $\beta$ decay; electron and muon capture}
\pacs{27.50.+e}{59 $\leq$ A $\leq$ 89}

\abstract{
The branching ratio between the $\beta^-$ and $\beta^+/\varepsilon$ decay of $^{74}$As has been measured in different host materials such as Ta, Al, Ge and mylar. No significant dependence of the branching ratio on the host material has been observed. The half-life of  $^{74}$As has also been measured in metallic Ta and in semiconductor Ge, no difference has been found and the results are in agreement with the literature value. The obtained results provide an upper limit for the possible host material dependence of the decay rate in $^{74}$As.
}

\begin{document}

\maketitle

\section{Introduction}

The low-energy cross section of charged particle fusion reactions is strongly affected by the electron screening phenomenon \cite{ass87}. The electron cloud surrounding the target nucleus reduces the Coulomb repulsion between the reacting nuclei and enhances the reaction cross section. This enhancement can be characterized by the $U_e$ screening potential. Recent systematic studies of the d(d,p)t reaction revealed that the screening potential becomes orders of magnitude higher (compared to the gas target value) when the target nuclei are in a metallic environment \cite{yuk98,cze01,rai02}. Presently, there is no generally accepted theoretical explanation to this strongly enhanced electron screening. Calculations based on the so-called dielectric function method \cite{cze04} could explain the observed target material dependence of the screening energy (at least for some of the experimental data) but it failed to reproduce its absolute value. The classical plasma theory of Debye has also been suggested as a possible explanation, where the equations of the plasma model are applied to the quasi-free metallic electrons \cite{rai04}. The model led to a reasonably good qualitative description of the measured screening potentials including the observed temperature dependence \cite{rai05}, however, the applicability of this model to the case of metallic environments is questionable (see below).

There is another proposed consequence of the enhanced electron screening which concerns the decay half-life of radioactive nuclides in a metallic environment. Similar to the case of charged particle induced reactions, the screening electrons may affect the emission of charged particles from a decaying nucleus. The emission of positively charged particles ($\alpha$ or $\beta^+$) can be enhanced by the electron cloud resulting in a reduced half-life, 
while longer half-life is predicted for negatively charged particle ($\beta^-$) emission\footnote{As it was shown recently by \cite{zin07}, however, the modified decay Q value due to the extention of the screening potential to the interior of the nucleus may partly cancel the effect of decay rate change.}. In the case of electron capture decay the prediction can be ambiguous: on one hand the plasma electrons may enhance the decay rate and on the other hand, the interaction between the bound electrons of the decaying nucleus and the high electron affinity host material atoms may result in a reduced decay rate. 

Different screening theories predict strikingly different decay rate changes for radioactive nuclei in metallic environment. While the dielectric function theory predicts only a modest change of decay rate and very weak temperature dependence \cite{cze06}, if one applies the Debye model, stronger decay rate changes are predicted and the effect is expected to be largely temperature-dependent since the Debye radius of the plasma electrons around the decaying nucleus depends on the temperature \cite{rai05}. Although the Debye model seems to describe well the observed enhanced electron screening in metallic environments, from theoretical point of view the applicability of the model to the system of metallic electrons is highly questionable since it requires that the temperature is larger than the Fermi temperature, i.e. typically above 10$^5$\,K. (see e.g. \cite{huk08}). However, since the idea gained a lot of publicity and it has also been suggested as a possible way towards the solution of nuclear waste disposal, huge experimental effort has been devoted to the study of its predictions. Besides a number of negative results \cite[e.g.]{nir07,sev07,sto07,goo07,rup08,kum08}, some measurements showed a slight change in the half-lives of various isotopes, although in some cases the effect was found to be only slightly larger than the experimental uncertainty \cite[e.g.]{lim06,wan06,spi07,rai07,jep07,gan08}. 

Up to now, the effect has been studied either by measuring directly the half-life of a given isotope in a certain environment (via the measurement of the activity as a function of time) or by measuring the activity of a sample under different conditions (e.g. at different temperatures). In both cases systematic errors may hinder the observation of small changes in the half-life or lead to false positive observation. In the case of direct half-life determination such a systematic error may come from e.g. not precisely known dead time of the counting system or from the long term drift in the detection efficiency. For the repeated activity measurement the not perfectly reproduced counting geometry may cause a systematic error. In the present work, a more precise method is presented that substantially reduces the systematic errors.

\section{The decay of $^{74}$A\lowercase{s}}

\begin{figure}
 \centering
  \includegraphics[angle=-90,width=0.9\columnwidth]{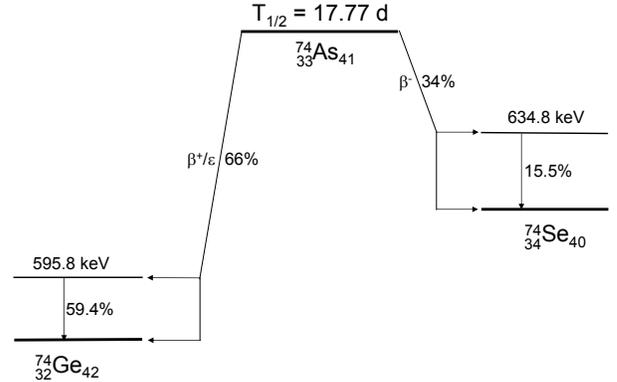}
 \caption{\label{fig:decayscheme} Simplified decay scheme of $^{74}$As showing the relevant $\gamma$-transitions. Weak decay branches to higher lying states in $^{74}$Ge and $^{74}$Se are not indicated \cite{sin06}.}
\end{figure}

In the chart of nuclides there are about two dozens of $\beta$-unstable nuclei for which it is energetically possible to undergo either a $\beta^-$ or a $\beta^+$ decay. $^{74}$As is one of these isotopes having also special advantages. On one hand, it decays with electron emission into $^{74}$Se with a branching ratio of $I_{\beta^-}$\,=\,34\,$\pm$\,2\,\%. On the other hand, it decays to $^{74}$Ge by positron emission  and electron capture ($I_{\beta^+/\varepsilon}$\,=\,66\,$\pm$\,2\,\%). Both decays are associated by strong $\gamma$-radiation from the de-excitation of the decay daughters. The $\beta^-$ decay is followed by an $E_\gamma$\,=\,634.8\,keV radiation (15.5\,$\pm$\,1.2\,\% relative intensity) and the $\beta^+/\varepsilon$ decay by an $E_\gamma$\,=\,595.8\,keV radiation (59.4\,$\pm$\,3.5\,\% relative intensity). In this latter case 26\% and 33.4\% relative intensities belong to the $\beta^+$ and electron capture decays, respectively. All decay data are taken from \cite{sin06}; Fig.\,\ref{fig:decayscheme} shows the simplified decay scheme of $^{74}$As. 

The electron screening should have an opposite effect on the $\beta^-$ and $\beta^+$ decays, therefore the measurement of the intensity ratio of the above two $\gamma$-radiations should be a sensitive probe of the screening effect on the decay. In spite of the apparent high interest towards the application of the Debye model to the half-life dependence on the host material, there is no detailed description of how to calculate the expected changes in the half-life of a given isotope with this model. Here, the strongly simplified calculation method of \cite{lim06} will be applied to the case of $^{74}$As without criticizing the method. The numbers will be given for the case of $^{74}$As in Tantalum, similar calculations can be done for Al (the other studied metal in the present work).

According to Ref. \cite{lim06} the $U_e$ electron screening energy in a metal can be written as

\begin{equation}
	U_e = 2.09\times10^{-11}Z(n_{eff}\rho_a/T)^{1/2}
\end{equation}
where $Z$ is the atomic number of the decaying nucleus, $n_{eff}$ is the number of conduction electrons per metallic atom, $\rho_a$ is the atomic density and $T$ is the temperature. For the d(d,p)t reaction the screening energy in Ta has been measured by \cite{rai04} and found to be 270\,$\pm$\,30\,eV at room temperature. Scaling this value to the case of electron or positron emmission of $^{74}$As ($Z$\,=\,33) one obtaines roughly 9\,keV. Similar number can be obtained by using directly the above formula with the known $n_{eff}$ value of Ta. Ref. \cite{lim06} suggests that the enhancement of the decay rate can be calculated as

\begin{equation}
	f \approx \left(\frac{Q+U_e}{Q}\right)^5
\end{equation}
where $Q$ is the Q-value of the $\beta$-decay to a given state. The $E_\gamma$\,=\,634.8\,keV radiation follows the $\beta^-$ decay of $^{74}$As to the first excited state in $^{74}$Se. This decay has a Q-value of 718\,keV. Substituting this to the above formula one obtaines a 6\,\% reduction of the $\beta^-$ decay rate of $^{74}$As to the given state. Similarly, the Q-value of the $\beta^+$ decay of $^{74}$As to the first excited state in $^{74}$Ge is 1967\,keV which leads to a 2.3\,\% higher $\beta^+$ decay rate. Taking into account the branching ratios of the different decays and assuming that the electron capture decay is not affected by the electron screening, the 595.8\,keV/634.8\,keV $\gamma$-intensity ratio in Tantalum should be about 7.5\,\% higher than in an insulator where no contribution from electron capture is expected.

The aim of the present work is to study this prediction by measuring the ratio of the 595.8\,keV and 634.8\,keV radiation intensities in the decay of $^{74}$As implanted into different materials. Metallic Ta, Al, semiconductor Ge and insulator mylar have been studied. Since the half-life of $^{74}$As (17.77\,$\pm$\,0.02 days) is short enough to be directly measured by following its decay, in addition to the high precision $\gamma$-intensity ratio measurements the half-life of $^{74}$As  has also been measured in the case of Ta and Ge samples albeit with inferior precision compared to the literature value.

\section{Experimental procedure}

%\subsection{Source preparation} 

\begin{figure}
 \centering
  \includegraphics[angle=-90,width=0.9\columnwidth]{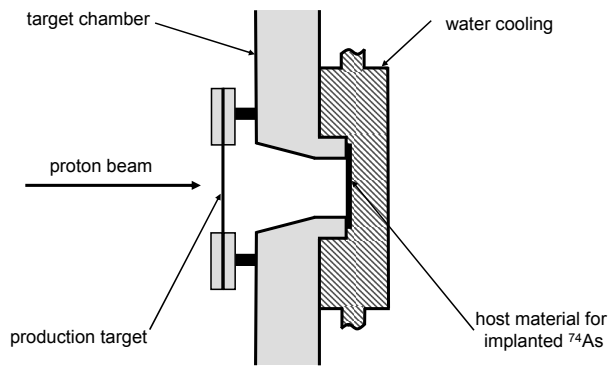}
 \caption{\label{fig:setup} Schematic view of the setup used for the $^{74}$As source production.} 
\end{figure}

The $^{74}$As sources have been produced by the $^{74}$Ge(p,n)$^{74}$As reaction in a setup schematically shown in Fig.\,\ref{fig:setup}. A 10.2\,MeV proton beam with 1\,$\mu$A intensity from the cyclotron of the ATOMKI bombarded a thin Ge target. This target was prepared by evaporating natural Ge onto a 3\,$\mu$m thick Al foil. The thickness of the Ge layer was roughly 50\,$\mu$g/cm$^2$. The target has been bombarded from the side of the Al foil backing. According to the reaction kinematics, the produced $^{74}$As nuclei emerge from the $^{74}$Ge(p,n)$^{74}$As reaction in a cone with about 55$^\circ$ maximum recoil angle and their energy varies between zero and 450 keV. The low-energy component of the recoiling $^{74}$As nuclei is stopped in the Ge layer and the high-energy part is implanted into the host material placed behind the production target. The implanted surface was 15\,mm in diameter. The host material served also as the beam-stop and was directly water cooled. 

Two metallic samples (Al and Ta) with 0.5\,mm thickness were used as host materials. Although the samples were directly bombarded by the proton beam, we observed no disturbing $\gamma$-activity. The evaporated Ge target itself was used to study the decay of $^{74}$As in Germanium. In this case, the size of the sample was estimated from the lateral beam size to be 8 \,mm in diameter. To implant  $^{74}$As into an insulator (which cannot be used as the beam-stop) the chamber has been extended behind the production target and the inner surface of the cylindrical extension was covered with an 8\,$\mu$m thick mylar foil. This foil served to catch the $^{74}$As nuclei emerging with a recoil angle between about 10$^\circ$ and 40$^\circ$. This corresponds to an implantation energy between roughly 220 and 440\,keV. After the implantation the mylar foil has been folded and fixed in a holder providing approximately the same source geometry as in the case of the metallic sources.

%\subsection{$\gamma$-counting}

Extensive simulations have been carried out to study the implantation profile of the produced $^{74}$As nuclei using the SRIM code \cite{SRIM}. In the simulations the reaction kinematics and the geometry of the production target and the host materials have been taken into account. The results show that more than 99.8\,\% of those nuclei which reach the host materials have energy higher than 10\,keV and are therefore implanted deep inside the crystal lattice of the studied samples.

The $\gamma$-activity of the samples has been measured with a 40\,\% relative efficiency HPGe detector. In order to minimize the coincidence summing effects, especially the summing of the 511\,keV annihilation radiation and the 595.8 keV radiation following the $\beta^+$-decay, a relatively large source-to-detector distance of 10 cm has been chosen. The whole counting setup was covered with lead shielding to reduce the laboratory background. From the known absolute efficiency of the detector, the source activities have been calculated and found to be in the range of a few hundred Bq for Ta, Al and mylar samples and 2 kBq for the Ge sample. Owing to the relatively low $^{74}$As activities and the absence of intense parasitic activities, the dead-time of the data acquisition system was always negligible.

\begin{figure}
 \centering
  \includegraphics[angle=-90,width=0.9\columnwidth]{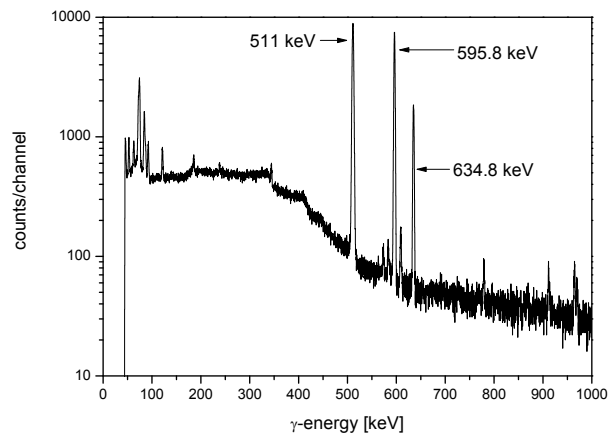}
 \caption{\label{fig:spectrum} Decay $\gamma$-spectrum of $^{74}$As implanted into mylar. The spectrum was taken for 100000 seconds three weeks after the source preparation. The two major peaks from $^{74}$As along with the 511\,keV annihilation peak are indicated.} 
\end{figure}

The energies of the two measured $\gamma$-rays are very close to each other, hence any change in the $\gamma$-detection efficiency (including changes in the counting geometry) is expected to have only a minor effect on the measured ratio. To investigate this effect, GEANT4 \cite{GEANT} simulations have been carried out for the 595.8\,keV/634.8\,keV efficiency ratio. In the simulations, different source distributions from the point-like case to 2 cm diameter have been considered as well as up to 1 cm off-axis displacement of the sources. All simulations resulted in the same efficiency ratios within the precision of about 0.5\%. The effect was also studied experimentally by repositioning the sources in the counting setup several times. The branching ratio for a given sample was found to be constant within the statistical uncertainty. Figure\,\ref{fig:spectrum} shows a typical $\gamma$-spectrum measured on the mylar sample. 

\section{Results}

%\subsection{The 595.8\,keV/634.8\,keV intensity ratios}

\begin{figure}
 \centering
  \includegraphics[angle=-90,width=0.9\columnwidth]{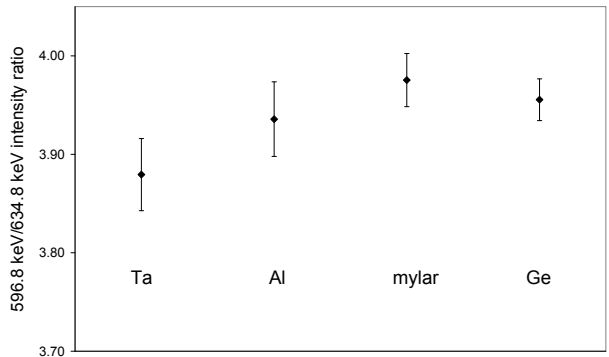}
 \caption{\label{fig:results} The obtained 595.8\,keV/634.8\,keV intensity ratios for the measured samples.} 
\end{figure}

The obtained 595.8\,keV/634.8\,keV intensity ratios for the four samples are shown in Fig.\,\ref{fig:results} and listed in Table\,\ref{tab:results}. The listed intensity ratios are corrected for the difference in the detector efficiency at the two energies ($\epsilon_{595.8\,\rm{keV}}/\epsilon_{634.8\,\rm{keV}}$\,=\,1.051\,$\pm$\,0.005, based on efficiency measurement with calibration sources), and for the 1.3\,$\pm$\,0.3\,\% summing out correction for the 595.8\,keV line obtained from GEANT4 simulations. The quoted uncertainties are quadratic sum of the statistical uncertainty and the 0.5\% systematic uncertainty from source geometry and positioning and do not contain the uncertainties from detector efficiency and summing correction which are common for all four measurements. The weighted average of the four samples gives an intensity ratio of 3.953\,$\pm$\,0.032 where the above common systematic uncertainties have been added quadratically. This is in agreement with the literature value of 3.83\,$\pm$\,0.37 which is calculated from the relative intensities of the 595.8\,keV and 634.8\,keV radiations (59.4\,$\pm$\,3.5\,\% and 15.5\,$\pm$\,1.2\,\%, respectively)\footnote{The veighted average of several direct measurements of the relative $\gamma$-intensities gives a more precise value of 3.852\,$\pm$\,0.082 for this ratio, again in good agreement with the present experiment.}.

\begin{table}
\caption{\label{tab:results} The obtained 595.8\,keV/634.8\,keV intensity ratios for the measured samples.}
\begin{tabular}{lccc}
\hline
\hline
Sample \hspace{2cm} &  \multicolumn{3}{c}{Intensity ratio} \\
\hline
Ta & 3.879 & $\pm$ & 0.037\\
Al & 3.936 & $\pm$ & 0.038\\
mylar & 3.975 & $\pm$ & 0.027\\
Ge & 3.955 & $\pm$ & 0.021\\
\hline
\hline
\end{tabular}
\end{table}

The results show a slight indication that the decay branching ratio is smaller for metallic samples (especially for Ta) compared with the insulator samples. However, the assumption of no dependence on the host material is well within two sigma errors of the measured points.

%\subsection{Direct half-life measurement of $^{74}$As}

\begin{figure}
 \centering
  \includegraphics[angle=-90,width=0.9\columnwidth]{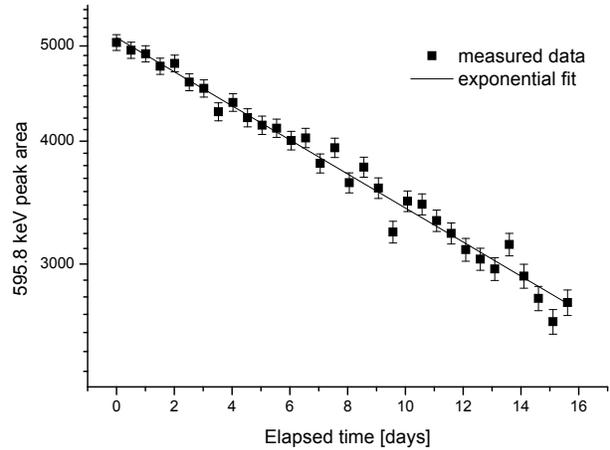}
 \caption{\label{fig:decay} Decay of $^{74}$As in Ta measured with the detection of the 595.8\,keV $\gamma$-ray. The reduced chi-square value of this fit is 0.46 while the other fits can be characterized by chi-square values in the range between 0.4 and 1.3.}
\end{figure}

In addition to the measured 595.8\,keV/634.8\,keV intensity ratios, the half-life of $^{74}$As has been directly measured in the case of Ta and Ge samples. The decay of the sources has been followed for 16 days (Ta sample) and for 13 days (Ge sample) recording the $\gamma$-spectra in every 12 hours (Ta) and 6 hours (Ge). In order to control any change or long-term variation in the detector efficiency, a $^{60}$Co source has been fixed to the Ge and Ta samples and the half-life has also been deduced by normalizing the counting rate from $^{74}$As with the 1173 and 1332\,keV lines from $^{60}$Co. The half-lives have been determined from the parameters of the exponential fit to the measured data. As an example, Fig.~\ref{fig:decay} shows the decay curve based on the 595.8 keV radiation for the Ta sample without normalization. In the case of the normalized half-life the decay of $^{60}$Co during the counting has been taken into account. From the known diffusion coefficient of As in Ge \cite{lan98} and in Ta \cite{lan90} one can conclude that at room temperature the diffusion of As is completely negligible, does not cause a loss of activity from the sample and therefore does not affect the half-life determination.

Table~\ref{tab:halflife} shows the measured half-life values for the two sources. The quoted values are the weighted averages of the ones obtained from the  detection of the 595.8\,keV and 634.8\,keV lines. The values obtained with and without normalization are in agreement confirming the long-term stability of the counting setup. The half-lives obtained without normalization are taken as adopted values of the present work, however, with an increased uncertainty typical for the normalized cases. The higher precision of the not-normalized half-life of the Ge sample is owing to the higher source activity (and therefore better statistics).

The half-lives for the Ta and Ge samples are in agreement with each other and with the literature value. This observation also supports that there is no host material dependence on the decay of $^{74}$As. 

\begin{table}
\caption{\label{tab:halflife} The measured half-life of the two samples. The quoted values are the weighted averages of the ones obtained from the  detection of the 595.8\,keV and 634.8\,keV lines.}
\begin{tabular*}{\columnwidth}{@{\extracolsep{\fill}}lcc}
\hline
\hline
$^{74}$As half-life [days] & Ta sample & Ge sample \\
\hline
without normalization &  17.87 $\pm$ 0.22 & 17.82 $\pm$ 0.04\\
normalized to 1173\,keV & 17.90  $\pm$ 0.24  & 17.94 $\pm$ 0.14\\
normalized to 1332\,keV &  17.87 $\pm$ 0.24 & 17.65 $\pm$ 0.16\\
adopted &  17.87 $\pm$ 0.24  & 17.82 $\pm$ 0.14\\
literature value \cite{sin06} &   \multicolumn{2}{c}{17.77 $\pm$ 0.02 } \\
\hline
\hline
\end{tabular*}
\end{table}

\section{Conclusions}

The results of the present work indicate that at room temperature for the studied four samples the intensity ratio of the 595.8\,keV and 634.8\,keV $\gamma$-rays in $^{74}$As decay does not depend on the material in which the decaying nuclei are embedded (the results are statistically not inconsistent with an up to 4\% difference between e.g. the Ta and mylar samples). From this observation consequences can be drawn concerning the possible changes of the $\beta^-$, $\beta^+$ and electron capture decay rates of $^{74}$As. The electron screening models predict a higher 595.8\,keV/634.8\,keV intensity ratio for metallic samples if one considers only $\beta^-$ and $\beta^+$ decays. This prediction is in contradiction with the observation of no host material dependence and especially with the measured slightly lower 595.8\,keV/634.8\,keV for the Ta sample compared with mylar. If, however, the electron capture decay rate is strongly reduced by the contribution of the plasma electrons, the effect may compensate (or overcome) the contribution from the $\beta^-$ and $\beta^+$ decays and lead to a reduction of the 595.8\,keV/634.8\,keV intensity ratio in metallic environment. This possibility can, however, be excluded based on the observed constant half-life using the following argumentation:

The decay rate of $^{74}$As can be written as the sum of the partial decay rates of the $\beta^-$, $\beta^+$ and electron capture decays:

\begin{equation}
\label{eq:halflife}
	\lambda \equiv \lambda_{\beta^-} + \lambda_{\beta^+} + \lambda_\varepsilon = {\rm constant}
\end{equation}
This value is constant within 2\% based on the present direct half-life measurement of the Ta and Ge samples. The 595.8\,keV/634.8\,keV intensity ratio measures the following quantity:
\begin{equation}
	\frac{a_{\beta^+}\cdot\lambda_{\beta^+} + a_\varepsilon\cdot\lambda_\varepsilon}{a_{\beta^-}\cdot\lambda_{\beta^-}} = {\rm constant}
\end{equation}
This value has also been measured to be constant within at most 4\%. Here, the $a$ coefficients stand for the emission probability of the measured $\gamma$-rays following the corresponding decay. Their values are the following \cite{sin06}: $a_{\beta^-}$ = 0.45, $a_{\beta^+}$ = 0.90 and $a_\varepsilon$ = 0.90. Owing to the fact that the $a$ coefficients for the $\beta^+$ and electron capture decays are equal, the latter formula can be written as:

\begin{equation}	
\label{eq:ratio}
	\frac{\lambda_{\beta^+} + \lambda_\varepsilon}{\lambda_{\beta^-}} = {\rm constant}
\end{equation}

Equations \ref{eq:halflife} and \ref{eq:ratio} can be satisfied simultaneously only if $\lambda_{\beta^-}$ is constant within about 3\% (the uncertainty here is based on the uncertainties of the measured half-lives and $\gamma$-intensity ratios using error propagation). This value is in clear contradiction with the calculated effect of  $\sim$7.5\%  from \cite{lim06}. The above considerations would still allow an equal and opposite change in the $\beta^+$ and electron capture decay rates, but this possibility is quite unlikely since the Debye model predicts similar effect on the $\beta^+$ and $\beta^-$ decays. Therefore, the conclusion of the present experiment is that at room temperature the rates of all three decays of $^{74}$As are unaffected by the host material within 3\% precision.

From the results of the present experiment (similar to the findings of some recent works) one can conclude that the application of the Debye model to the case of radioactive decay in metallic environment can be clearly excluded, at least the magnitude of change of the decay rate predicted by \cite{lim06} cannot be reproduced by the experiment for the studied isotope. Although the result of the present work is in agreement with a null dependence of the decay rate on the host material, the modest decay rate change suggested by the dielectric function theory \cite{cze04} is statistically not inconsistent with the results. Since the two theoretical approaches predict strikingly different temperature dependences of the decay rate, the study of the decay of $^{74}$As at low temperature would strengthen further this conclusion. Such an experiment in the mK temperature range is in progress.

\acknowledgments
This work was supported by the Economic Competitiveness Operative Programme GVOP-3.2.1.-2004-04-0402/3.0. and by OTKA (K68801, T49245). Gy. Gy. acknowledges support from the Bolyai grant.

\end{document}